\documentclass[aps,preprint,superscriptaddress]{revtex4}
\usepackage{amsfonts,amssymb}
\usepackage{graphicx,pstricks}

\begin{document}


\title{Downhill versus two--state protein folding in a statistical
  mechanical model}


\author{Pierpaolo Bruscolini}
\email{pier@unizar.es}
\affiliation{Instituto BIFI, Universidad de Zaragoza, c.\ Corona de
  Arag\'{o}n 42, Zaragoza, Spain}

\author{Alessandro Pelizzola}
\email{alessandro.pelizzola@polito.it}
\affiliation{Dipartimento di Fisica and CNISM, Politecnico di Torino,
  c. Duca degli Abruzzi 24, Torino, Italy}
\affiliation{INFN, Sezione di Torino}

\author{Marco Zamparo}
\email{marco.zamparo@polito.it}
\affiliation{Dipartimento di Fisica and CNISM, Politecnico di Torino,
  c. Duca degli Abruzzi 24, Torino, Italy}


\begin{abstract}
We address the problem of downhill protein folding in the framework of
a simple statistical mechanical model, which allows an exact solution
for the equilibrium and a semi--analytical treatment of the
kinetics. Focusing on protein 1BBL, a candidate for downhill folding
behavior, and comparing it to the WW-domain of protein PIN1, a  two--state folder of
comparable size, we show that there are qualitative differences in
both the equilibrium and kinetic properties of the two
molecules. However, the barrierless scenario which would be expected
if 1BBL were a true downhill folder, is observed only at low enough
temperature.
\end{abstract}


\maketitle


\section{Introduction}
\label{sec:intro}

The folding  of small single--domain proteins is often described as a
two--state process, where a molecule starts from, e.g., a disordered
thermodynamic state and evolves towards its ordered (native)
thermodynamic state with a kinetics characterized by a single
exponential behavior. If a suitable reaction coordinate can be
identified, and the corresponding free energy profile can be computed,
one expects, in the vicinity of the denaturation temperature, a
profile with two minima separated by a barrier.

In recent years it has however been suggested \cite{MunozBBL1} 
that there are proteins whose kinetics, at all temperature, 
might not be hindered by a free energy barrier, 
thus reproducing the  
 ``downhill folding'' scenario suggested in \cite{Downhill}.
Typical features of a downhill folder would be a continuous variation of the
thermodynamic state and nonexponential time behavior. Protein 1BBL
has been thoroughly investigated by Mu\~{n}oz and coworkers
\cite{MunozBBL1,MunozBBL2,MunozBBL3,MunozBBL4,MunozBBL5,MunozBBL6,MunozBBL7,Munoz2007nat}
and several results seem to indicate that it is one such downhill
folder, though they have been questioned by other authors
\cite{FershtBBL1,FershtBBL2,FershtBBL3,Fersht2007PNAS,Fersht2007nat,Zhou2007nat}.

Purpose of the present paper is to investigate the folding behavior
of 1BBL, compared to the WW-domain of protein PIN1, a clear two--state folder of comparable
size, in the framework of a simple statistical mechanical model, which
allows an exact solution of the equilibrium and a semi--analytical
approach to the kinetics. We are going to consider the
Wako--Sait\^{o} 
(WS) model
\cite{WS1,WS2,ME1,ME2,ME3,Amos1,Amos2,ItohSasai1,ItohSasai2,HenryEaton,ItohSasai3,AbeWako},
a one--dimensional model with binary degrees of freedom and
long--range, many--body interactions, which has found applications
even outside the protein folding domain \cite{TD1,TD2,TD3}. The WS
model has already been applied to this problem in \cite{MunozBBL1},
where an approximate solution for the equilibrium was used. In the
present paper we shall make use of the exact solution for the
equilibrium, following the approach described in \cite{BP,P}, while
for the study of the kinetics we shall resort to Monte Carlo
simulations and the local equilibrium approach \cite{ZP-PRL,ZP-JSTAT}.

The plan of the paper is as follows: in Sec.\ \ref{sec:model} we shall
describe the WS model, then the equilibrium behavior of the BBL and
PIN1 molecules will be studied in Sec.\ \ref{sec:eq} after a careful
discussion about the choice of model parameters. Kinetics will be
discussed in Sec.\ \ref{sec:kin}, and in Sec.\ \ref{sec:land} we shall
analyze the molecule's behavior from the perspective of free energy
profiles. Finally, our conclusions will be drawn in Sec.\
\ref{sec:concl}.

\section{The model}
\label{sec:model}

The WS model was introduced in 1978 by Wako and Sait\^{o} \cite{WS1,WS2}. Yet, it was largely unknown to the wide community of researchers in protein folding when, about two decades later, it was independently reintroduced by 
Eaton and coworkers
\cite{ME1,ME2,ME3,HenryEaton}, with minor differences with respect to the former model, as a simple and efficient theoretical tool to interpret their experimental data. Thanks also to this experiment-oriented approach, the model achieved some popularity and was then used by many different authors for
several purposes
\cite{Amos1,Amos2,ItohSasai1,ItohSasai2,ItohSasai3,AbeWako}, even
in a problem of strained epitaxy \cite{TD1,TD2,TD3}. In the following we
shall use our exact solution of the equilibrium thermodynamics
\cite{BP,P} and our local equilibrium approach (LEA) to the kinetics
\cite{ZP-PRL,ZP-JSTAT}. 

The model is a G\={o}--like model \cite{Go} which considers a protein as a
sequence of $N+1$ aminoacids connected by C--N peptide bonds. The
degrees of freedom of the model are associated to the $N$ peptide
bonds and will be denoted by $m_k$, $k = 1,2, \ldots N$. These are
binary variables which take values in $\{0,1\}$. In the case $m_k = 1$
the $k$--th peptide bonds is assumed to be in a native--like (or
ordered) conformation, while $m_k = 0$ stands for an unfolded (or
disordered) conformation. For each peptide bond the set of unfolded
conformations is of course larger than the native one, and this is
taken into account in an effective way by introducing the entropy cost
$q_k > 0$ of ordering bond $k$. In \cite{IPZ} it is shown how to give
an explicit realization of this entropy cost in terms of microscopic
degrees of freedom. The main feature of the model is that two
aminoacids interact only if they are in contact in the native state
(non--native interactions are neglected, in the spirit of G\={o}--like
models) and if all the peptide bonds between them are in the native
state (that is, the corresponding dihedral angles $\phi$, $\psi$
assume their native values). The latter is a drastic assumption which
makes the model amenable to analytic treatments, up to the exact
solution of the equilibrium. Nevertheless, the model has been shown to
give realistic results for the kinetics of protein folding.

The effective free energy (sometimes called "effective Hamiltonian" in the physics
literature) of the model can be written as
\begin{equation}
H = \sum_{i=1}^{N-1}\sum_{j=i+1}^{N}
\epsilon_{i,j} \Delta_{i,j} \prod_{k=i}^{j} m_k
- R T \sum_{k=1}^{N} q_k (1-m_k),
\label{Hamiltonian}
\end{equation}
where $R$ is the gas constant and $T$ the absolute
temperature. $\Delta$ is the contact matrix, and its $(i,j)$ element
takes value 1 if aminoacids $i$ and $j+1$ are in contact in the native
state (that is, if they have at least a pair of atoms
closer than 0.4 nm according to the structure deposited in the Protein
Data Bank \cite{pdb}) and 0 otherwise. The corresponding contact
energy $\epsilon_{i,j} < 0$ is defined as in \cite{ME3} as $- k
\epsilon$ if the number $n_{\rm at}$ of atom pairs in
contacts satisfies $5(k-1) < n_{\rm at} \le 5k$. Here $\epsilon$ is an
energy scale which is determined, together with the entropic costs
$q_k$, as described in the next section. 

\section{Equilibrium}
\label{sec:eq}

The first exact solution of the equilibrium was already reported in
\cite{WS1,WS2} and then went forgotten until, as far as we know, the
original papers were cited again in \cite{ItohSasai1}. In the
meanwhile, we had developed our approach \cite{BP,P} to this exact
solution, which we shall follow in this paper. It relies on a mapping
to a two--dimensional problem, which stems from the introduction of
the new binary variables $x_{i,j} = \displaystyle{\prod_{k=i}^j} m_k$
for $1 \le i \le j \le N$, which take value 1 only if all the peptide
bonds belonging to the stretch from $i$ to $j$ are in the native
state, and 0 otherwise. In the case $i = j$ we have of course $x_{i,i}
= m_i$. These new variables are apparently non--interacting, since the
effective free energy Eq.\ (\ref{Hamiltonian}) is a linear function of
them, but they actually interact through the constraints $x_{i,j} =
x_{i+1,j} x_{i,j-1}$ which they must satisfy. Our new variables can be
associated to a triangle--shaped portion of a two--dimensional
(square) lattice and the model can be easily solved by transfer
matrix, since due to the constraints the matrices involved are at most
of rank $N$. In \cite{BP} it is shown how to compute the partition
function, the free energy as a function of the number of native
peptide bonds and relevant expectation values.

Before proceeding to describe the equilibrium properties of 1BBL, we
discuss the choice of the model parameters. Let us first of all
consider the contact map. Different model proteins 
have been
considered in the literature, corresponding to the same core sequence with different choices of N and C terminal parts.
For instance, the protein corresponding to  PDB code 1BBL has been used in \cite{MunozBBL1}, while 1W4H has been
used in \cite{FershtBBL1}. The corresponding sequences differ only in
the end parts, and have a 45 residue identical subsequence, which
goes from residue 7 to 51 for 1BBL and from 126 to 170 for 1W4H; the relative
native structures have disordered terminal parts. 
Since the theoretical model we use requires the knowledge of the atomic coordinates of the protein in the native structure, we 
consider the longest common subsequence with resolved structure, and take residues 12
to 48 in 1BBL , together with their atomic coordinates from
the file 1BBL.pdb. 

We can now move on to the choice of the parameters $\epsilon$ and
$q_k$. 
The entropic costs $q_k$ have often been
chosen as in \cite{ME3,BP}, and take values $q_H$ for the more
structured parts of the molecule (the peptide bonds preceeding a
residue marked by B, E, G, H, I or T according to the DSSP
classification \cite{DSSP}), and $q_C$ for the remaining, less
structured parts. To begin with we take $q_H = 1.66$ and $q_C = 0.6$
as in \cite{ME3}.
	
\begin{figure}
\centerline{\includegraphics*[width=8cm]{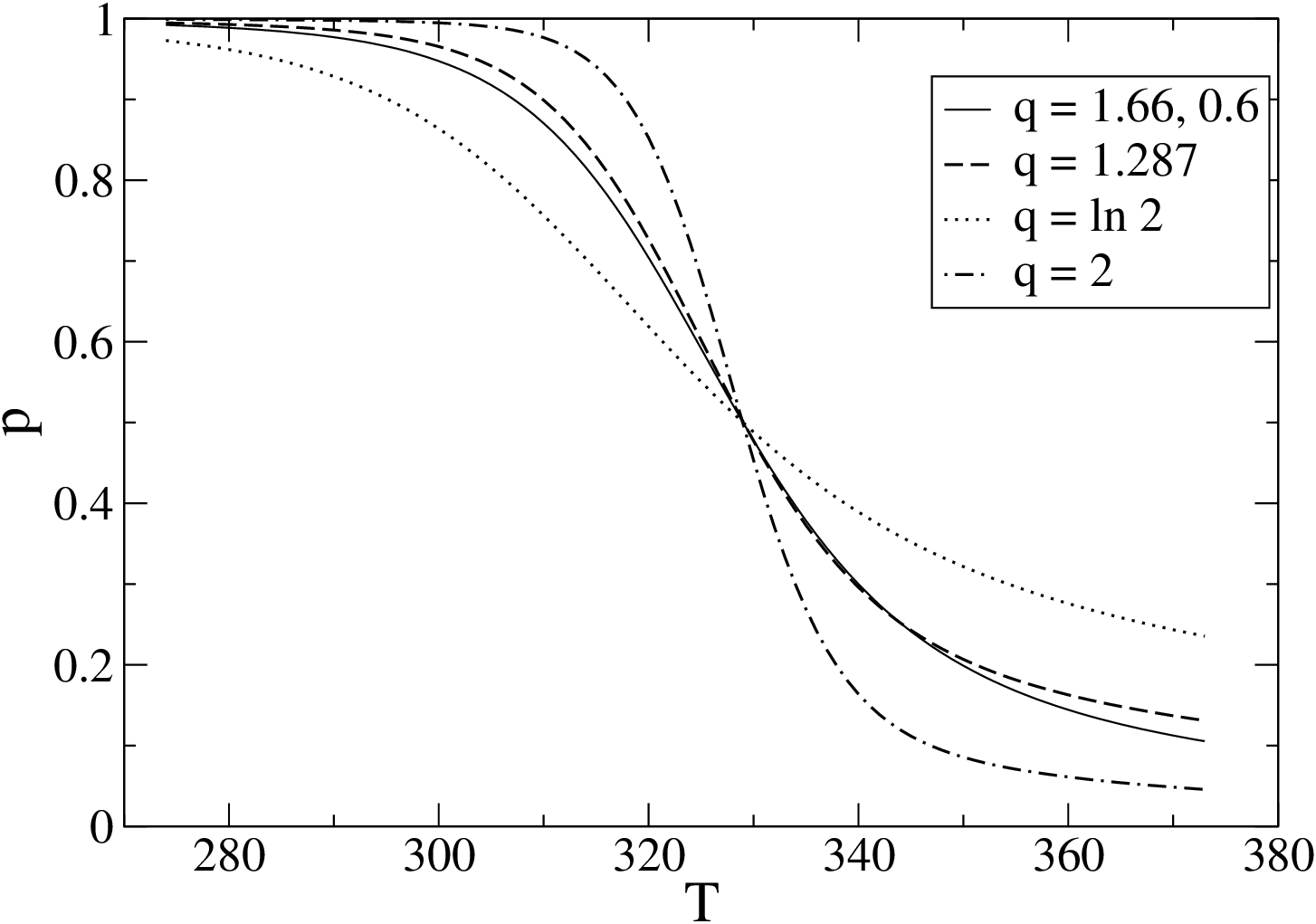}}
\caption{\label{fig:MvsT} Average fraction of folded molecules as a
function of temperature (in Kelvin). 
See text for the discussion of the meaning of the various
lines.}
\end{figure}

The energy scale $\epsilon$ is then determined by imposing the
condition that at the experimental denaturation temperature the
fraction $p$ of folded molecules is 1/2. Several estimates of the
denaturation temperature have been reported. To begin with, we
consider $T_m = 329$ K, which is consistent with the estimates in
\cite{FershtBBL1} (this choice will be refined in the following). In
order to give an estimate of $p$ we introduce the number of native
peptide bonds $M = \sum_{k=1}^N m_k$ and the average fraction $m =
\displaystyle{\frac{\langle M \rangle}{N}}$ of such bonds. $m$ takes
the value 1 at zero temperature and $m_\infty =
\displaystyle\frac{1}{N}\displaystyle\sum_{k=1}^N (1 + e^{q_k})^{-1}$
at infinite temperature. A good definition of the fraction of folded
molecules is then $p = \displaystyle\frac{m - m_\infty}{m_0 -
m_\infty}$, where $m_0$ is a value which represents well the folded
state. For 1BBL we can choose $m_0 = m(T=0) = 1$. The energy scale
$\epsilon$ has therefore to satisfy $p(T_m) = 1/2$. In this way we
obtain $\epsilon/R = 99.8$ K and the temperature dependence of $p$ is
shown in Fig.\ \ref{fig:MvsT} (solid line).

In order to simplify our parameter choice we can try to remove the
distinction between more and less structured parts of the molecule
(notice that the heterogeneity of the system is not removed, it
remains encoded in $\Delta_{ij}$) and, keeping $\epsilon$ fixed to the
above value, look for a unique value $q_k = q$, $k = 1, 2, \cdots N$
such that $p(T_m) = 1/2$. We obtain $q = 1.287$ and the dashed line in
Fig.\ \ref{fig:MvsT}. Since we do not aim at reproducing in detail the experimental results, but just to grasp the basic features of the folding of 1BBL with a minimal model,  we consider that the changes introduced by the simplification are negligible  and from now on we consider a single parameter $q$.

\begin{figure}
\centerline{\includegraphics*[width=8cm]{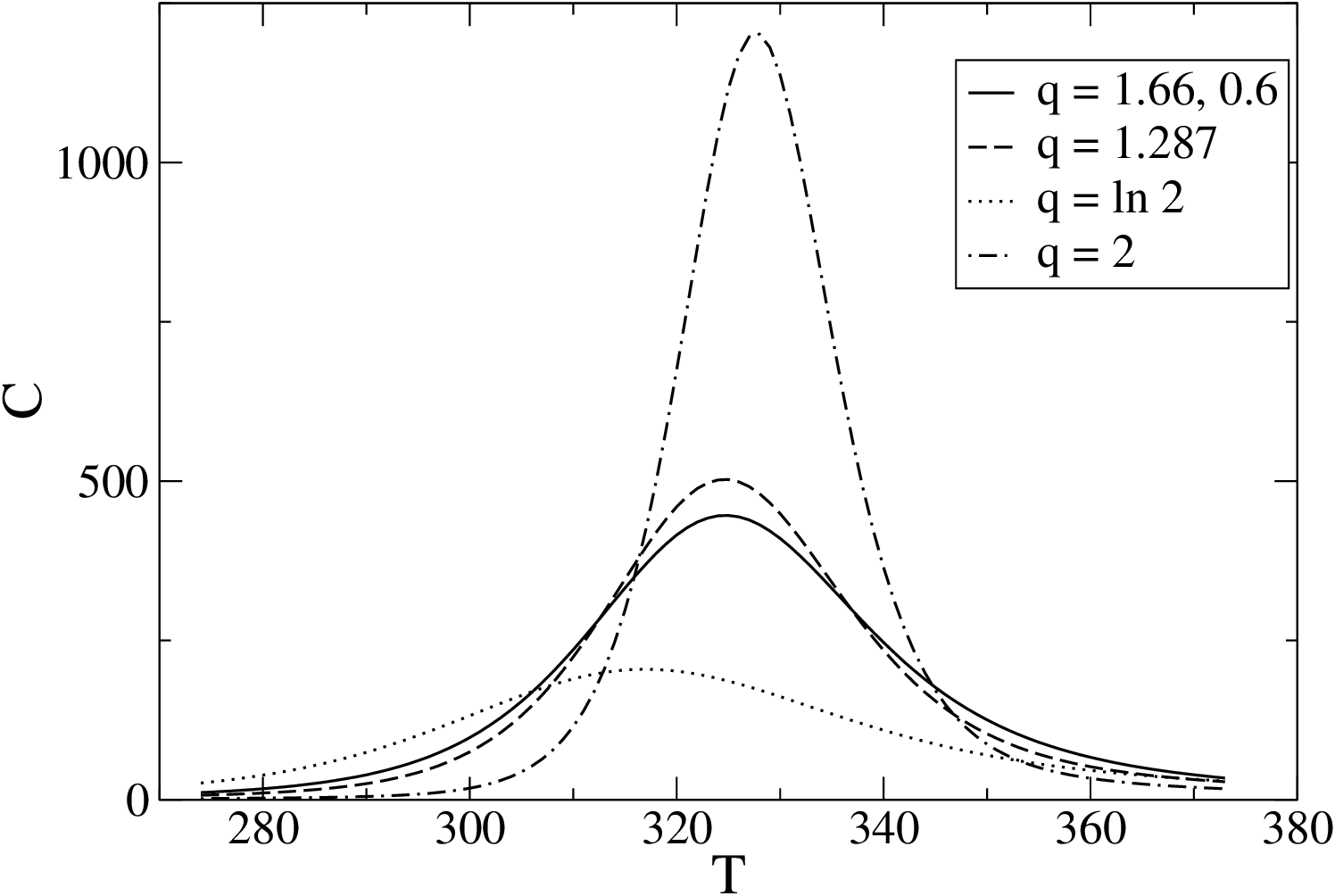}}
\caption{\label{fig:CvsT} 
  Specific heat  for protein 1BBL, vs temperature, for the same values of $q$ as in Fig.\ \ref{fig:MvsT}. T is reported in Kelvin, specific heat $C$ in Kcal/ (K mole), so that the plotted quantity is adimensional. }
\end{figure}

The effect of $q$ on the temperature dependence of $p$ is clearly seen
in Fig.\ \ref{fig:MvsT} where two other values have been considered,
and $\epsilon$ has been adjusted every time so that $p(T_m) =
1/2$. Choosing $q = \ln 2$, as in \cite{IPZ}, we have $\epsilon/R = 73.0$ K and the dotted line, while
for $q = 2$, as previously chosen for a model antiparallel $\beta$--sheet \cite{ZP-PRL},
we obtain $\epsilon/R = 137$ K and the dash--dotted line. It is
clearly evident that $\epsilon$ and, as a consequence, the sharpness
of the transition, increase with $q$. The reinforcement of the
transition for increasing $q$ (and $\epsilon$) is also seen in Fig.\
\ref{fig:CvsT}, where the temperature dependence of the specific heat
is shown.

\begin{figure}
\centerline{\includegraphics*[width=8cm]{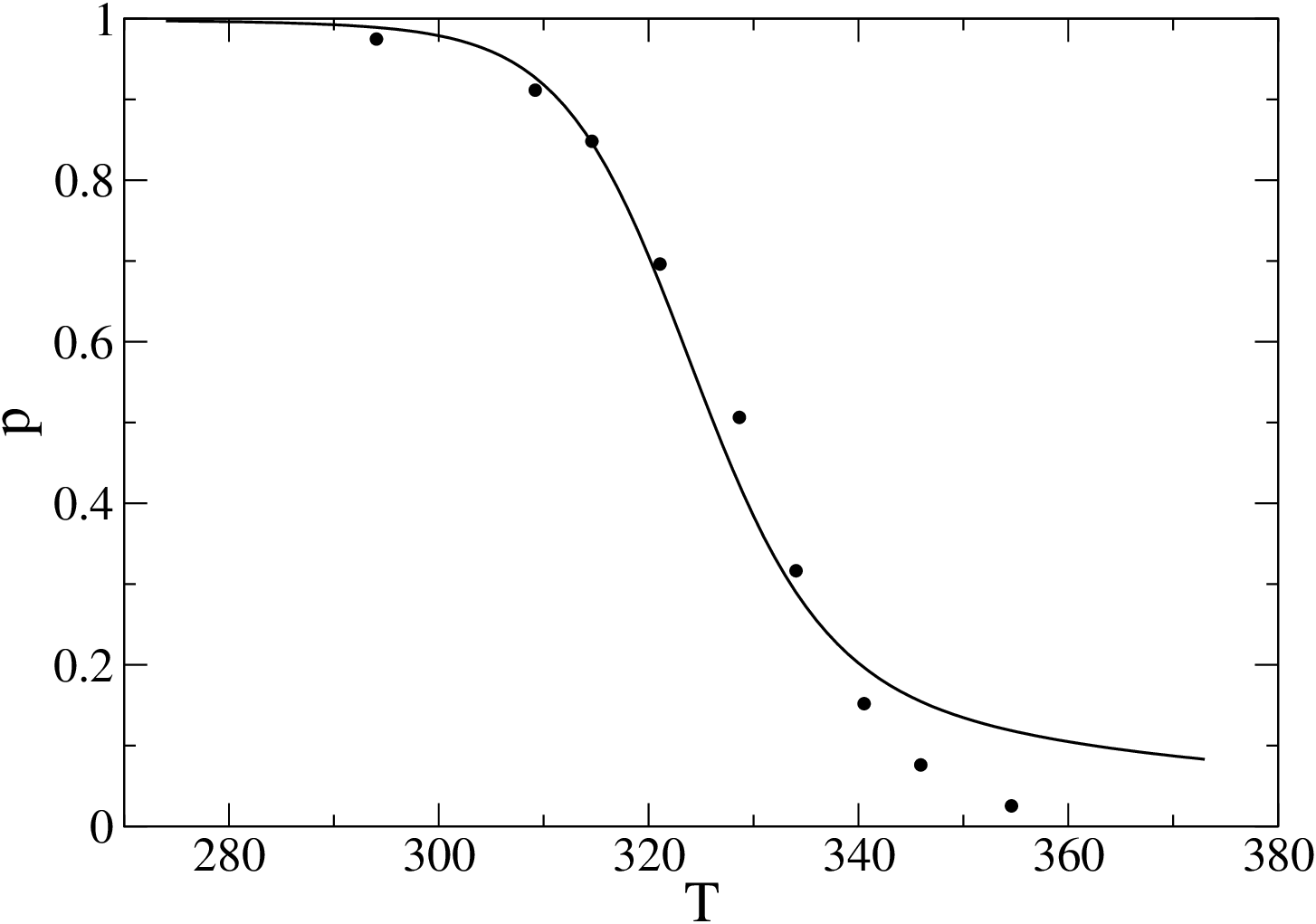}}
\caption{\label{fig:pFit} Average fraction of folded molecules as a
function of temperature. Fit to experimental data.}
\end{figure}

In order to determine a pair $(\epsilon,q)$ for our analysis we try to
fit the experimental results for the native fraction as a function of temperature (from circular dichroism and NMR) in \cite{FershtBBL1}, Fig.\ 2(b). 
We infer the set of data from the coordinates reported in the figure itself, and fit them with theoretical curves by our model, obtaining, as an estimate for our parameters, the values $q = 1.589$ and $\epsilon/R
= 114$ K, which we shall use from now on. These correspond to $T_m =
326.2$ K, which is consistent with the estimates reported in
\cite{FershtBBL1}. 
The fit 
is reported in Fig.\ \ref{fig:pFit}.  The quality of the fit is not optimal, especially at high temperatures, where apparently there is a change in the unfolded minimum as T increases. 
 It could probably be improved by introducing enough heterogeneity in the interaction and entropy parameters $\epsilon_{i,j}$ and $q_i$, but this would introduce a huge number of parameters to be fitted, in the end at the expenses of simplicity and of a clear understanding of the basic features underlying the folding of 1BBL. 
Alternatively,
we have  checked  (data not shown) that improving quality at high temperature by setting $\epsilon/R=140$ K and $q=2.075$ (to ensure the same value of $T_m$),  at the expenses of the low temperature region, introduce minor changes on the free-energy profiles reported in Section \ref{sec:land}. On the other hand, we have also noticed
that if we blindly assumed that at T=373 K the protein is completely denaturated, using  $m(T=373 K)$ instead of $m_\infty$ as the baseline for the denatured state in the definition of $p$, then it is possible to get a much better fit of the data, by only adjusting $q$ to  $q=1.580$: a small change that leaves the profiles (see Sec.\ \ref{sec:land}) almost unchanged. We mention this here to stress how the choice of the baselines can dramatically affect the quality of the fits; however, in the following we go back to our first choice of parameters, and to the original and correct choice of $m_\infty$ as the true baseline, just
%
noticing that Fig.\ \ref{fig:pFit} suggests that our model behaves in a less cooperative way than the true protein: we will take this into account when discussing our results for kinetics and profiles, in the next sections.

\begin{figure}
\centerline{\includegraphics*[width=8cm]{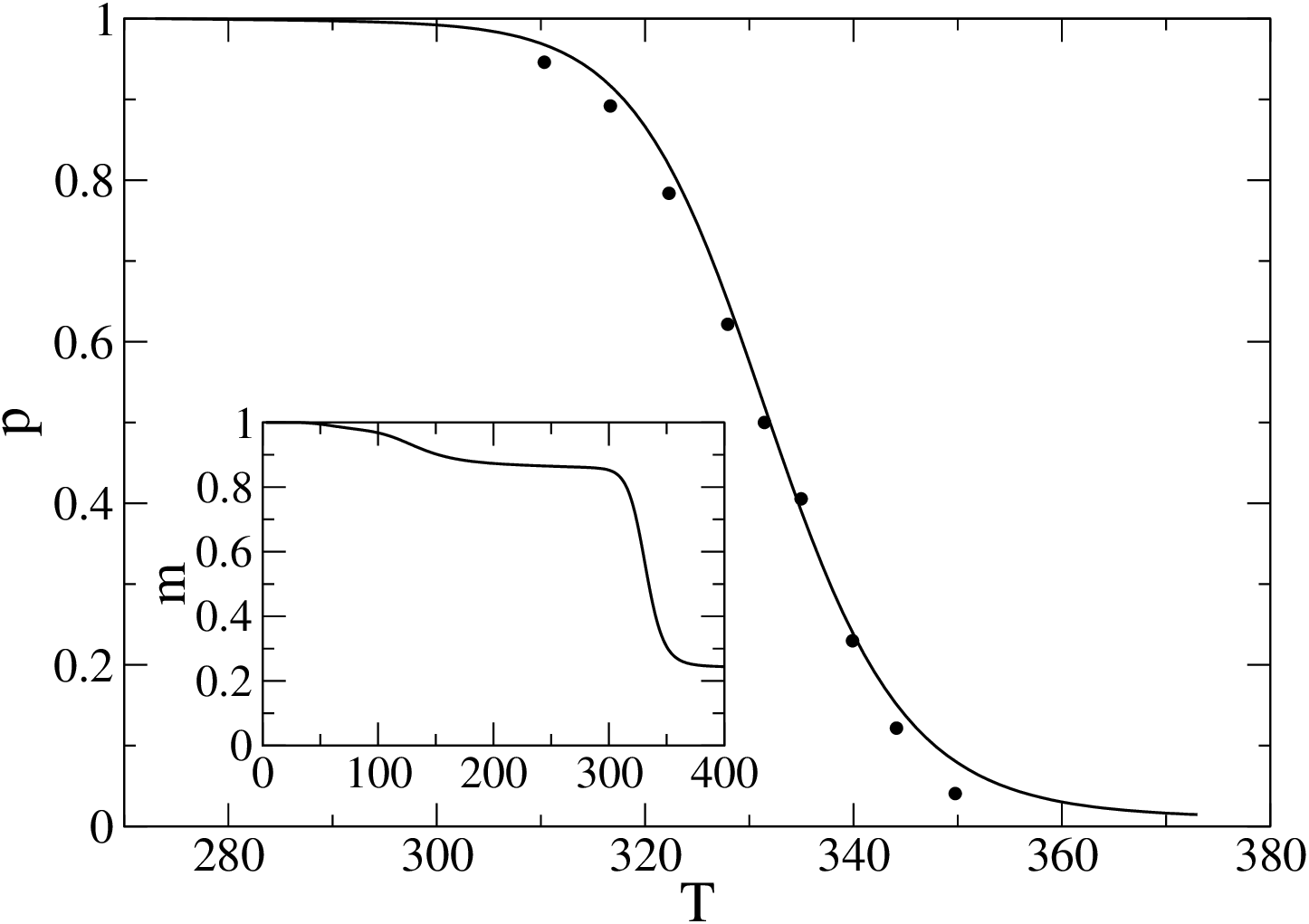}}
\caption{\label{fig:pFitPIN1} Same as Fig.\ \ref{fig:pFit} for protein
  the WW-domain of protein PIN1. Inset: plot of the fraction of native peptide units, $m$, over a wider temperature range. Notice that a short tail of the protein, corresponding to the residues having little structure in the native state, denature at very low temperature. We disregard this behavior and define $p$ to account for the jump around $T_m$=332 K, where the protein unfolds completely and cooperatively.}
\end{figure}

The same procedure above is carried out for the WW-domain of protein PIN1 (pdb code 1I6C; for simplicity in the following we will often refer to it as PIN1),
which is a clear two--state folder of a size very close to the 1BBL one
(38 peptide bonds instead of 36) and will be used for comparison
throughout the paper. By using $T_m = 332$ K and fitting to the data
in \cite{Gruebele}, Fig.\ 3(d), we obtain $q = 1.185$ and $\epsilon/R
= 60$ K. See Fig.\ \ref{fig:pFitPIN1} for the corresponding fit.
Here the reference value $m_0$ for the folded state has been chosen as
$m_0 = m(T = 273 K) < 1$, since within this model the molecule orders
perfectly only at temperatures $T < 50$ K, while a wide plateau in
$m(T)$ is observed in the range 200 to 300 K (see inset). 

From the above results, we can see a first difference between 1BBL and PIN1, in that the experimental results concerning the latter can be better fitted than those relative to the former, when the same level of description is used for the two systems. That is, as long as only the geometric aspects of the native states are considered, by imposing that $\epsilon_{i,j}=\epsilon$ and $q_i=q$ for all $i$ and $j$, the model performs better in reproducing the all-$\beta$ protein Pin1 WW-domain rather than the helical one 1BBL, and the latter appears to be less cooperative in the model than in reality.

We conclude this section by observing that in any case from the equilibrium
results it is difficult to extract reliable information about the
two--state vs.\ downhill behavior of the molecules, 
due to the fact that observables like the average fraction of native residues, i.e. those that correspond to the first moment of the probability distribution, cannot distinguish between unimodal and multimodal distributions, and can assume the same value  in the case the underlying distribution is that of a two-state folder as well as in the case of a downhill one. 
In principle, specific heat conveys more information, and a measure of cooperativity is represented by the ratio $\kappa$ of van't Hoff to calorimetric enthalpy (see Ref.\ \cite{Chan2000} for a critical discussion of several definition of this parameter). 
However, also in that case, it is not easy to put a clear threshold between two-state, cooperative behavior, that ideally corresponds to $\kappa=1$, and less cooperative one, for $\kappa<1$. 

To obtain a deeper insight on the folding mechanism of the two proteins, more
detailed studies of the kinetics and of the free energy profiles are
needed, and will be done in the next sections.

\section{Kinetics}
\label{sec:kin}

In the present section we shall study the time behavior of our model
when it is subject to a simple Metropolis kinetics (the effect of
different choices for the kinetics has been discussed in
\cite{ZP-PRL}). More precisely, we consider a single ``bond--flip''
kinetics, defined by the discrete--time master equation:
\begin{equation}
p_{t+1}(x) = \sum_{x'} W(x' \to x) p_t(x),
\label{Master}
\end{equation}
where $x = \{ x_{i,j}, 1 \le i \le j \le N \}$ denotes a configuration
which satisfies the constraints described in the previous section,
$p_t(x)$ is the probability of configuration $x$ at time $t$. The
transition probability $W(x \to x')$ is assumed to vanish if $x$ and
$x'$ differ by more than one peptide bond, is given by: 
\begin{equation}
W(x \to x') = \frac{1}{N} {\rm min} \left\{ 1, \exp \left[ -
  \frac{H(x') - H(x)}{RT} \right] \right\},
\end{equation}
if $x$ and $x'$ differ by exactly one peptide bond and $W(x \to x) = 1
- \displaystyle\sum_{x' \ne x} W(x \to x')$ by normalization. 

The kinetics can be studied by two different approaches, namely direct
Monte Carlo (MC) simulations and the LEA developed
in \cite{ZP-PRL,ZP-JSTAT}, where it is assumed that the system
probability evolves in a restricted space given by those probabilities
which satisfy the same factorization property as the equilibrium one
(these can be thought of as equilibrium probabilities of models with
the same WS structure but different values of the parameters). It
has been shown in \cite{ZP-PRL} that this approximation is quite
accurate for proteins of a size comparable to 1BBL.

\begin{figure}
\centerline{\includegraphics*[width=8cm]{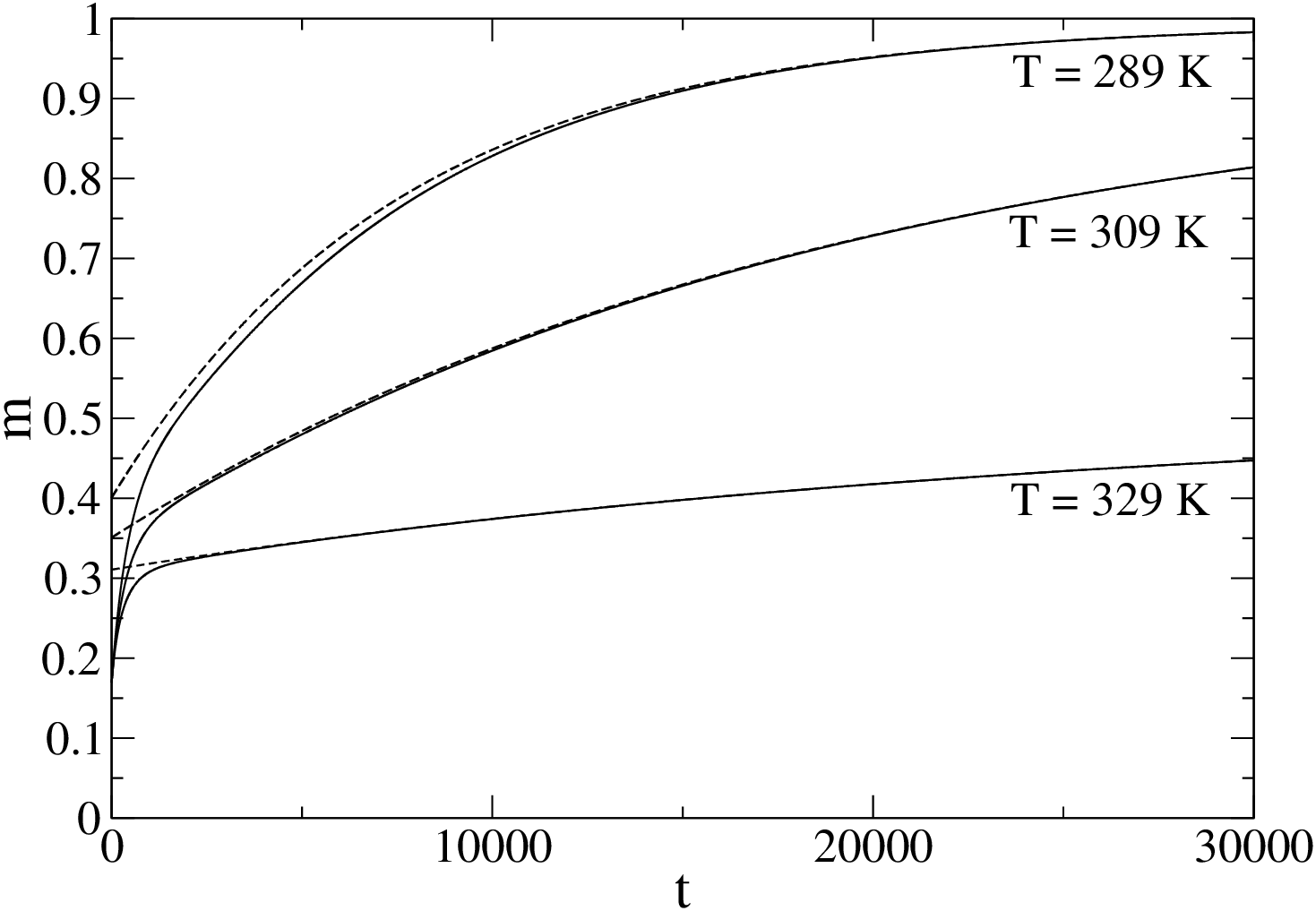}}
\caption{\label{fig:LEA} Average fraction of native peptide bonds vs.\
  time in LEA (solid lines) and single exponential fits (dashed
  lines) for protein 1BBL.}
\end{figure}

In Fig.\ \ref{fig:LEA} we report the average fraction of native bonds
$m$ as a function of time (solid line) at the temperature $T = 329$ K
and at two lower temperatures, 309 K and 289 K. The initial condition
is disordered, corresponding to equilibrium at infinite
temperature. Single exponential fits $m(t) = m_\infty - a
\exp(-t/\tau)$ are also reported and the equilibration times are $\tau
= 2.64 \times 10^4$, $1.95 \times 10^4$ and $7.56 \times 10^3$ for $T =
329$, 309 and 289 K respectively; time unit, here and in the following, is the elementary time corresponding to proposing a change of the state of a peptide bond. It is clearly seen that the
deviations from a single exponential behavior are more relevant for
smaller temperatures, that is when $\tau$ is smaller. Notice that the
$t$ axis extends over a duration slightly longer than the longest
$\tau$.

\begin{figure}
\centerline{\includegraphics*[width=8cm]{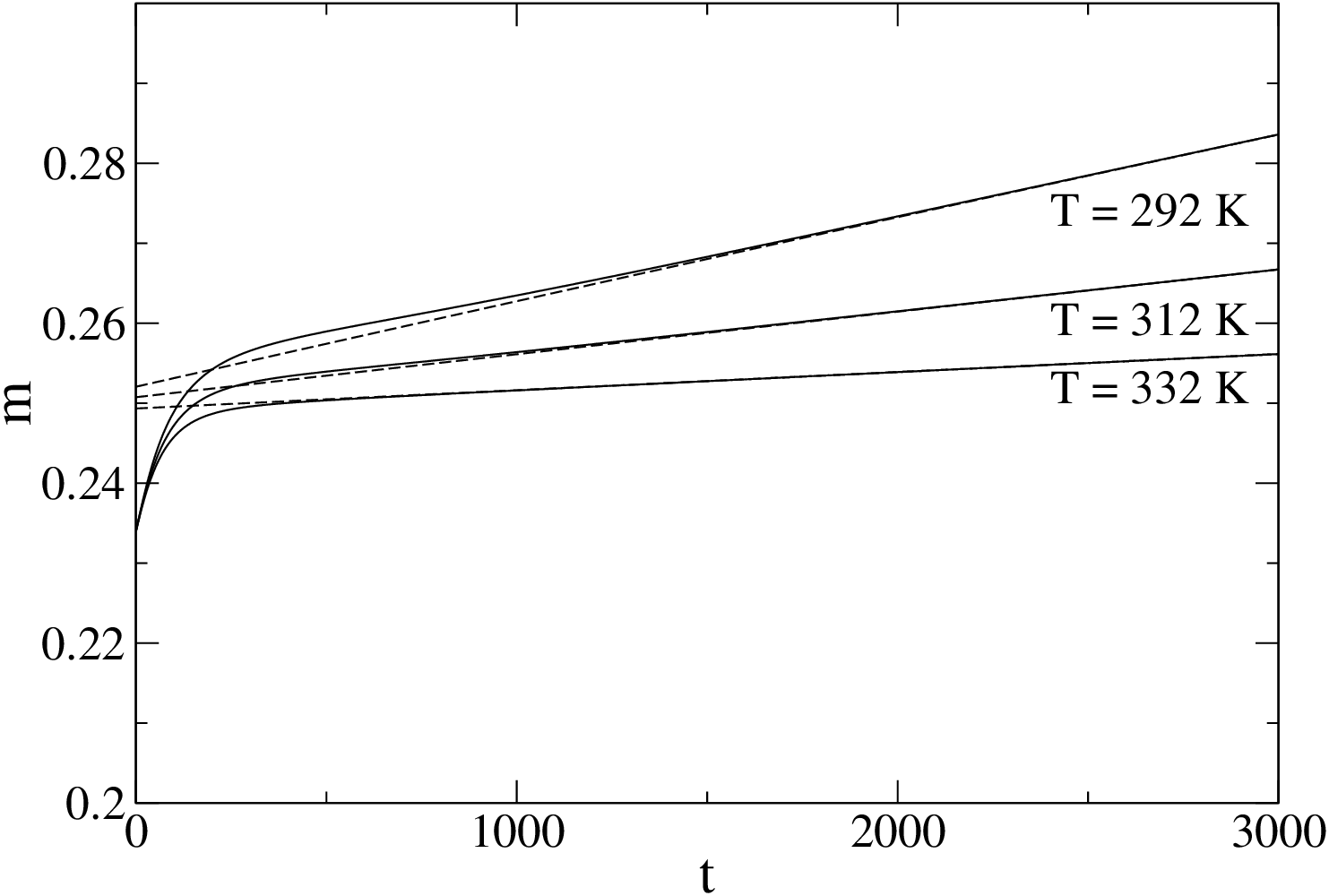}}
\caption{\label{fig:LEA-PIN1} Same as Fig.\ \ref{fig:LEA} for protein
PIN1.}
\end{figure}

For comparison, we have repeated the same analysis for protein PIN1,
and the corresponding results are reported in Fig.\
\ref{fig:LEA-PIN1}. We considered temperatures $T = $332, 312
and 292 K, the equilibration times being $\tau = 1.31 \times 10^5$,
$1.06 \times 10^5$ and $5.61 \times 10^4$ respectively. These
characteristic times have been computed, within the present model, in
\cite{ZP-PRL} for a wide range of temperatures, and turned out to
reproduce quite well the experimental data. Observe that 1BBL is
considerably faster than PIN1, roughly 5 times at the denaturation
temperature. Here the deviations from the single exponential
behavior are much less relevant (notice that here the $t$ axis
extends over a duration of order 1/40th of the largest $\tau$).

It has been suggested \cite{Downhill} that downhill folding
can imply a nonexponential time behavior. Although the large time
single exponential behavior is formally a direct consequence of Eq.\
\ref{Master} one might argue that the much larger deviations we
observe for 1BBL could be a signature of a potential downhill folding
behavior. Several types of analytical time behavior, as alternatives
to the single exponential, have been proposed for downhill folders,
including multiexponential 
and stretched
exponential \cite{
GruebeleDownhill}, which are actually
related to each other \cite{JohnstonStretched}. In order to test these
ideas we made multiexponential fits, using up to three exponentials,
to our results. 

The fits are made with the following procedure: the relaxation rates 
obtained previously from equilibrium (long time) analysis are factored out
from the kinetics data for $m(t)$, and a fit is performed according to 
\begin{equation}
\left(m(t) -m(\infty) \right) e^{k t} = a + b e^{-\mu t} ,
\label{eq:2expfit}
\end{equation}
or
\begin{equation}
\left(m(t) -m(\infty) \right) e^{k t} = a + b e^{-\omega t} + c e^{-\mu t},
\label{eq:3expfit}
\end{equation}
so that the second and third smallest rates are $k+\omega$ and $ k+\mu$.
Results for the two-exponential fit are reported on Table \ref{tab:ratefit}.

\begin{table}
\begin{center}
\begin{tabular}[b]{|r|r|r||r|r|r|}
\multicolumn{3}{c}{    1BBL}  &    \multicolumn{3}{c}{     WW-domain} \\
\hline 
T [K]  & k [1/$\tau_f$] & $\mu$[1/$\tau_f$] & T [K] & k[1/$\tau_f$] & $\mu$ [1/$\tau_f$]  \\
\hline 
289 & 1.3  $\times$ 10$^{-4}$ & 4.31  $\times$ 10$^{-5}$ & 
292 & 1.78 $\times$ 10$^{-5}$ & 1.55 $\times$ 10$^{-2}$ \\
309 & 5.08 $\times$ 10$^{-5}$ & 4.6  $\times$ 10$^{-5}$ & 
312 & 9.47 $\times$ 10$^{-6}$ & 1.46 $\times$ 10$^{-2}$ \\
329 & 3.78 $\times$ 10$^{-5}$ & 2.87  $\times$ 10$^{-3}$ & 
332 & 7.64 $\times$ 10$^{-6}$ & 1.36 $\times$ 10$^{-2}$ \\
%
\hline
\end{tabular} 
\end{center}
\caption{Results for the two-exponential fit of the rates for 1BBL and PIN1 WW-domain. See text for discussion. Here $\tau_f$ is the elementary time associated to the flip of the state of a peptide unit.}
\label{tab:ratefit}
\end{table}
%
%
%
%

From these we notice that the dominant correction to
the main exponential behavior decays with a characteristic time which
is of the same order (at low temperatures) or up to two orders of magnitude (at higher T) smaller than $\tau$ for 1BBL; on the contrary,  PIN1 WW-domain
 exhibits a time scale separation of 4--5 orders of magnitude. 
The above remarks hold true also for the three-exponential fits: in the case of 1BBL, the introduction of  $\omega$ induces an adjustment of $\mu$ with respect to the value previously found, but both $\omega$ and $\mu$ stay within the two-orders of magnitude range found above. For the PIN1 WW-domain one gets the same $\mu$ , while the third rate $k+\omega$ represents just a small correction of the relaxation rate $k$ (data not shown), and strongly correlates with it, signalling that the two-exponential fit is enough to grasp the essential features of the spectrum. 
The difference in the rates is a further confirmation that 1BBL deviates more than PIN1 WW-domain from the
single-exponential time-behavior: since the existence of a relevant barrier induces a separation of time scales, the small gap in the rates for 1BBL suggests that the barrier is not very pronounced in this case.
It is interesting to notice that the above findings are in agreement with the results for a different fast-folding protein, $\lambda_{6-85}$, studied in Ref.\  \cite{GruebeleNat2003}, where the observed departure from simply-exponential kinetics can be described by a Langevin dynamics on a suitable double well potential with a small barrier.

\begin{figure}
\centerline{\includegraphics*[width=8cm]{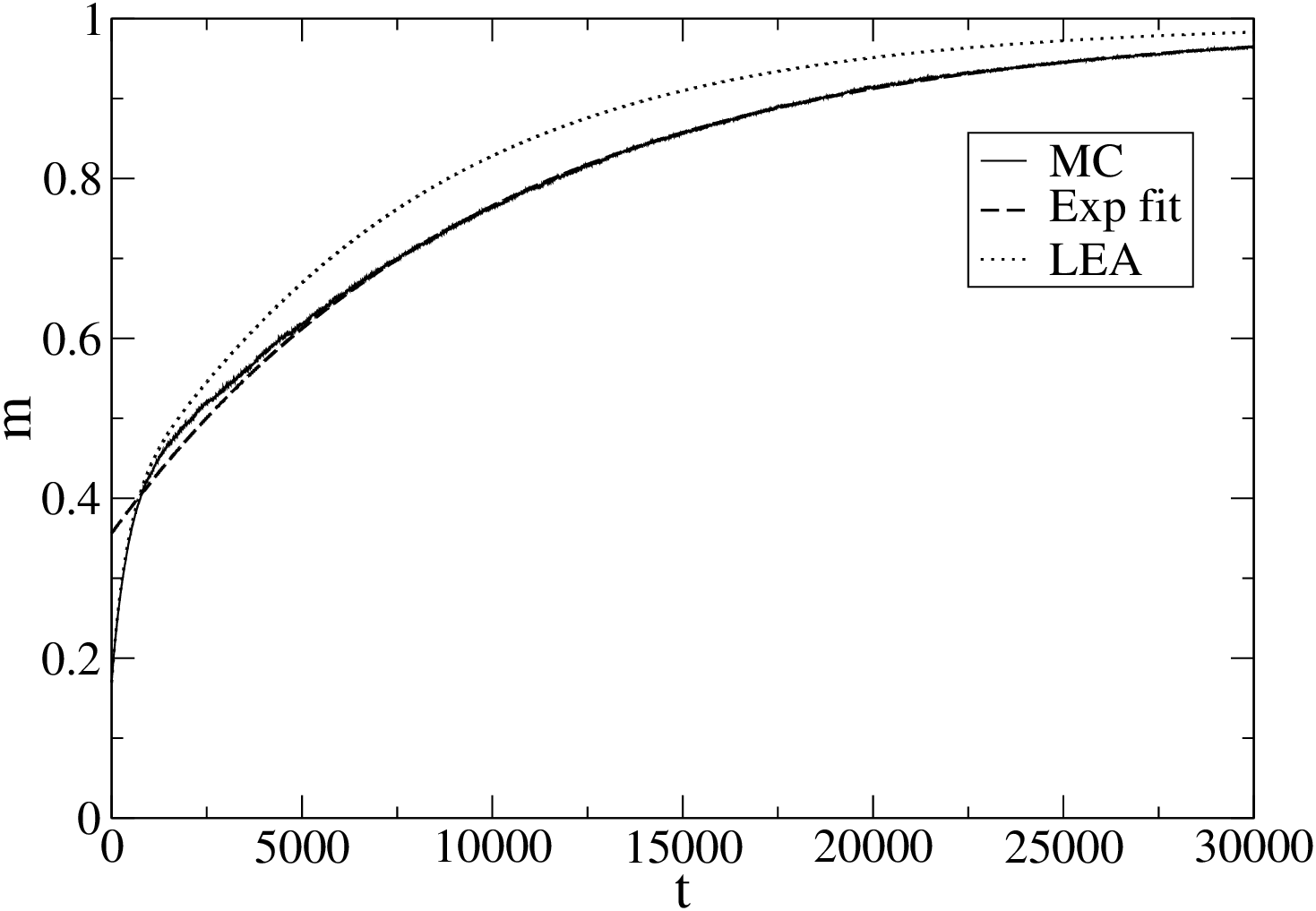}}
\caption{\label{fig:MCave} Average fraction of native bonds as a
  function of time for 1BBL at $T = 289$ K. Solid line: MC, dashed
  line: single exponential fit to MC, dotted line: LEA.}
\end{figure}

\begin{figure}
\centerline{\includegraphics*[width=8cm]{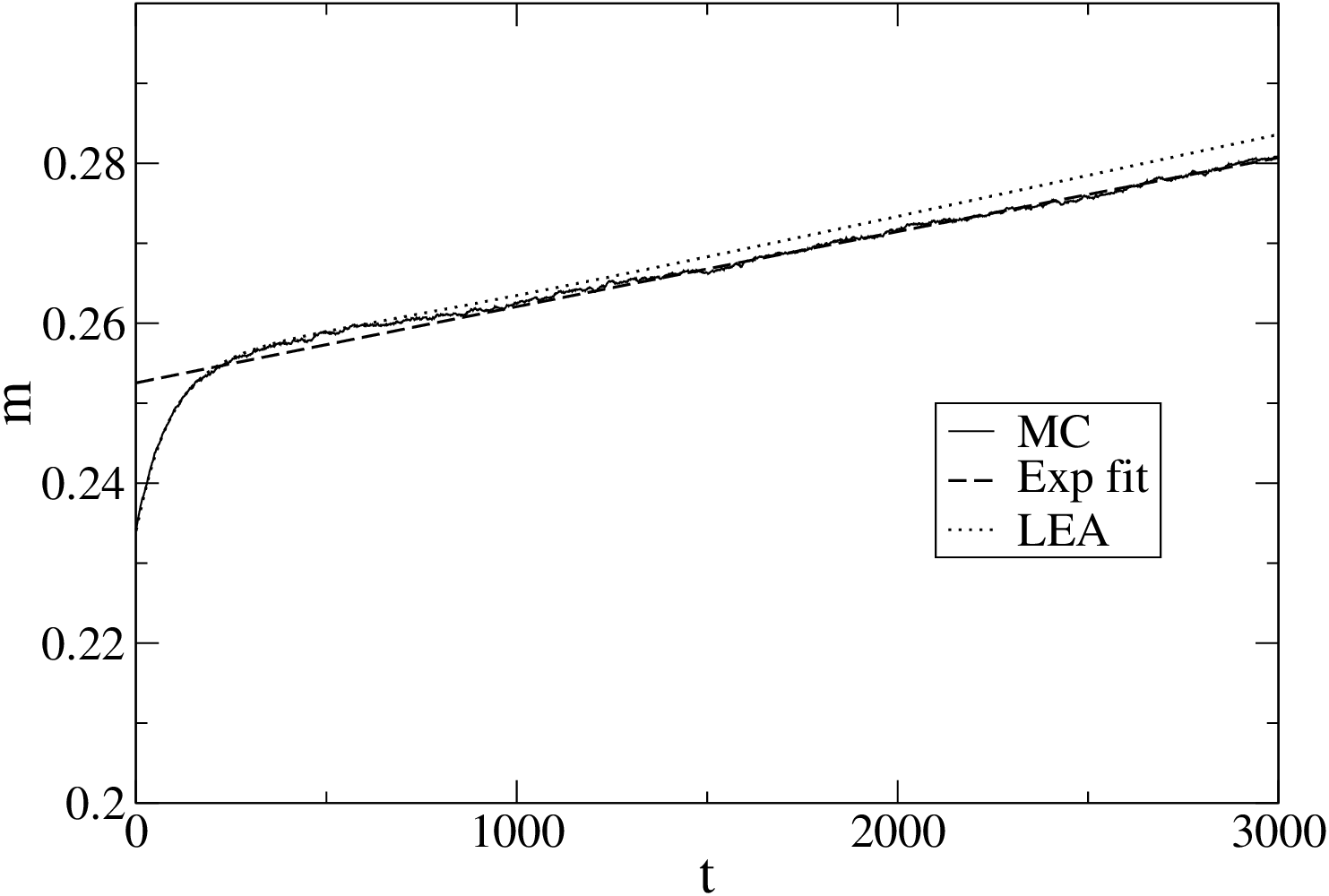}}
\caption{\label{fig:MCave-PIN1} Same as Fig.\ \ref{fig:MCave} for PIN1 at
  $T = 292$ K.}
\end{figure}

Before accepting the above arguments it is however important to check
that the above results are not artifacts of the local equilibrium
approximation. An exact solution of the kinetics is not feasible, but
MC simulation can provide a reference result if we average
over a large enough number of trajectories (which is much more time
consuming than LEA). We have therefore compared
the LEA results at the lowest temperatures, both for 1BBL and PIN1,
with MC results, averaging over $10^4$ trajectories for 1BBL
(Fig.\ \ref{fig:MCave}) and $10^5$ trajectories for PIN1 (Fig.\
\ref{fig:MCave-PIN1}).

One can see that our conclusions about the deviation from the single
exponential behavior are confirmed by MC simulations. In addition, we
recall that LEA characteristic times are lower bounds of the exact ones
\cite{ZP-PRL,ZP-JSTAT} and indeed our MC simulations give $\tau \simeq
9.74 \times 10^3$ for 1BBL and $\tau \simeq 6.30 \times 10^4$ for PIN1.

\begin{figure}
\centerline{\includegraphics*[width=8cm]{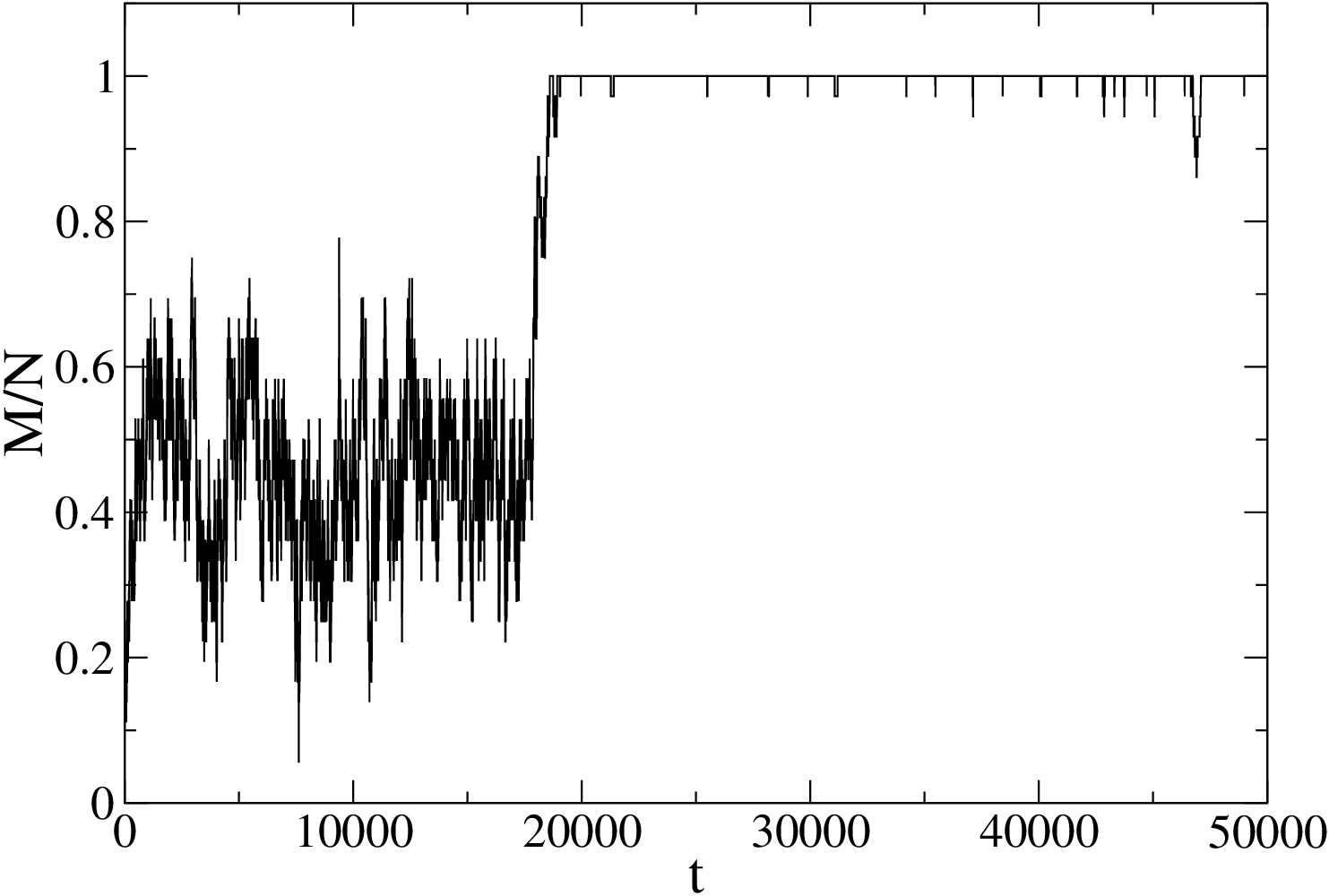}}
\caption{\label{fig:singleMC} Fraction of native peptide bonds vs.\
  time in a Monte Carlo simulation, starting from a disordered
  configuration, at $T = 289$ K.}
\end{figure}

It is also interesting to take a look to single MC trajectories, which
might be representative of the behavior of single molecules.  In
Fig.\ \ref{fig:singleMC} we report the time behavior of the fraction
$M/N$ of native peptide bonds in a typical Monte Carlo simulation at a
temperature $T = 289$ K, well below the denaturation temperature,
starting from a disordered state. It is clearly evident that the
behavior resembles that of a two--state system, with an almost
saturated ordered state and a rather broad disordered state,
characterized by large fluctuations. Similar results are obtained at
temperatures closer to $T_m$, where of course the ordered state
fluctuates more and the system switches repeatedly between the two
states.

\begin{figure}
\centerline{\includegraphics*[width=8cm]{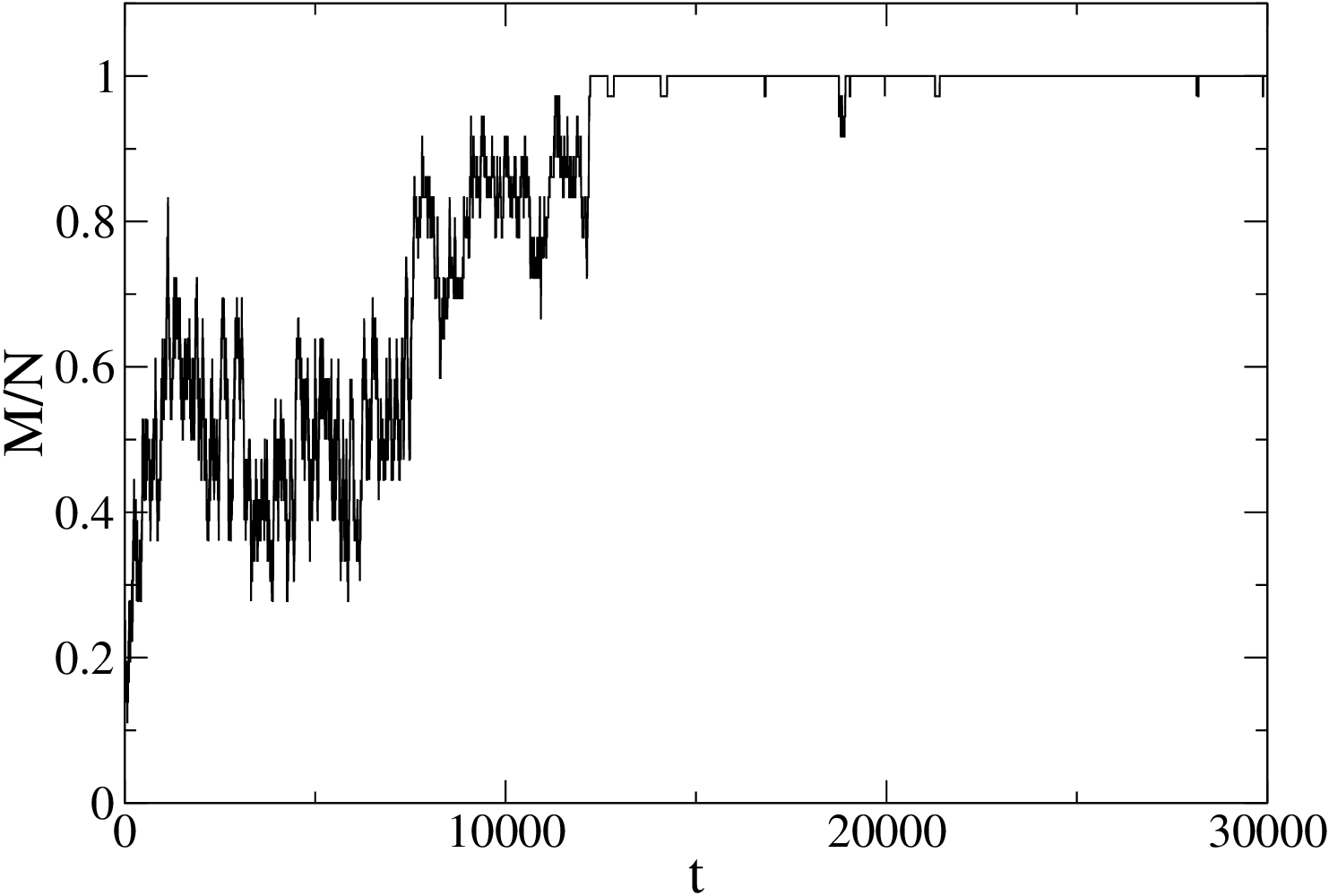}}
\caption{\label{fig:singleMCcold} Same as Fig.\ \ref{fig:singleMC}, at
$T = 269$ K.}
\end{figure}

In order to observe a different, continuous--like behavior, one has
to go to even smaller temperatures, for instance $T = 269$ K (Fig.\
\ref{fig:singleMCcold}). Notice that at the same temperature PIN1
still has a clear two--state behavior, which persists down to 170 K.

Aiming to get a deeper understanding of these results and to complete
our analysis of the 1BBL behavior, we move on to the study of its free
energy profile as a function of the number of native peptide bonds,
which will be the subject of the next section.

\section{Free energy profiles}
\label{sec:land}

\begin{figure}
\centerline{\includegraphics*[width=8cm]{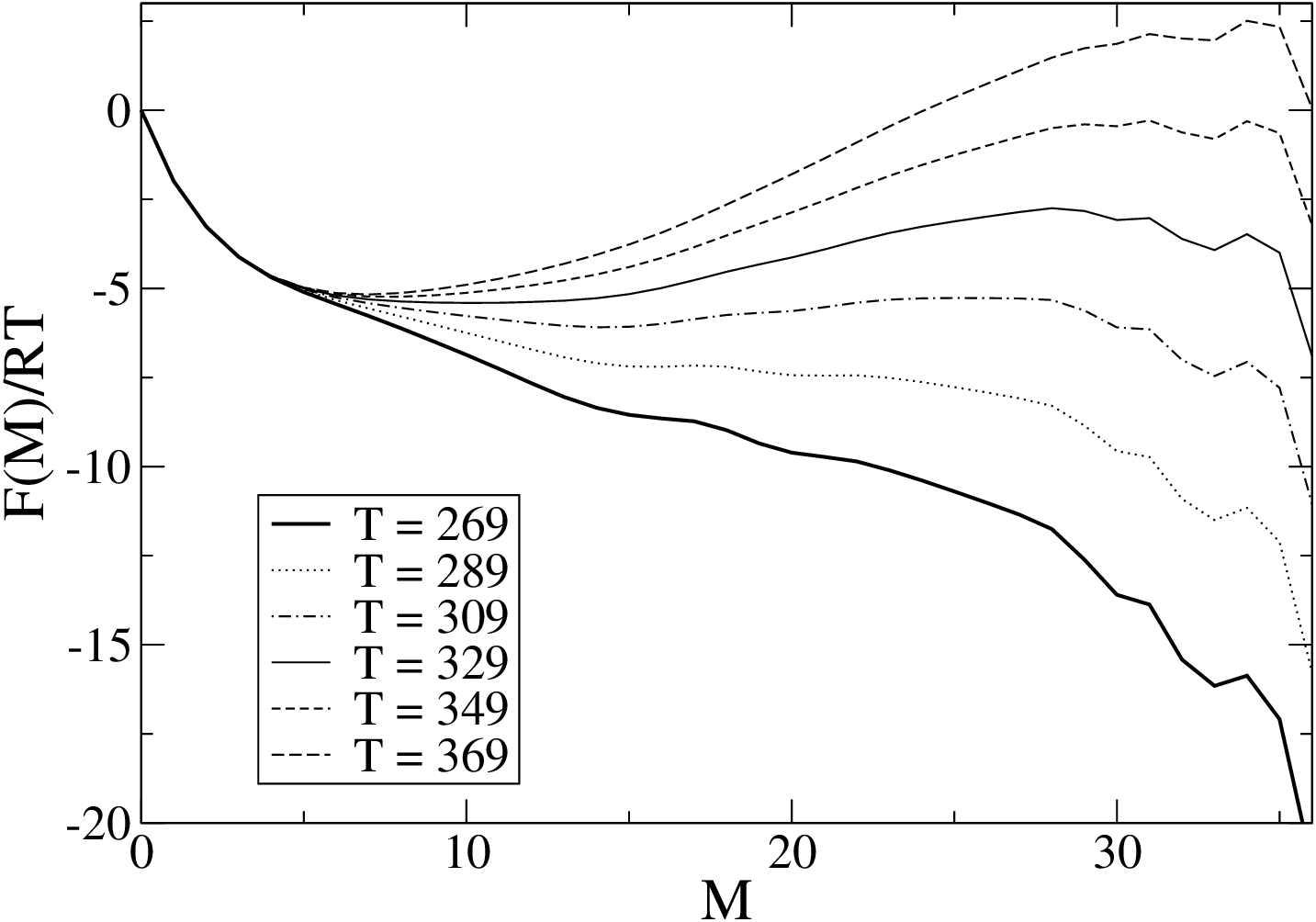}}
\caption{\label{fig:EqLand} The 1BBL free energy profile
  $F(M)$ for various temperatures.}
\end{figure}

The equilibrium free energy profile $F(M)$ as a function of the
number $M$ of native peptide bonds can be easily computed in an exact
way as shown in \cite{BP}, and the result is reported in Fig.\
\ref{fig:EqLand} for various temperatures. 

\begin{figure}
\centerline{\includegraphics*[width=8cm]{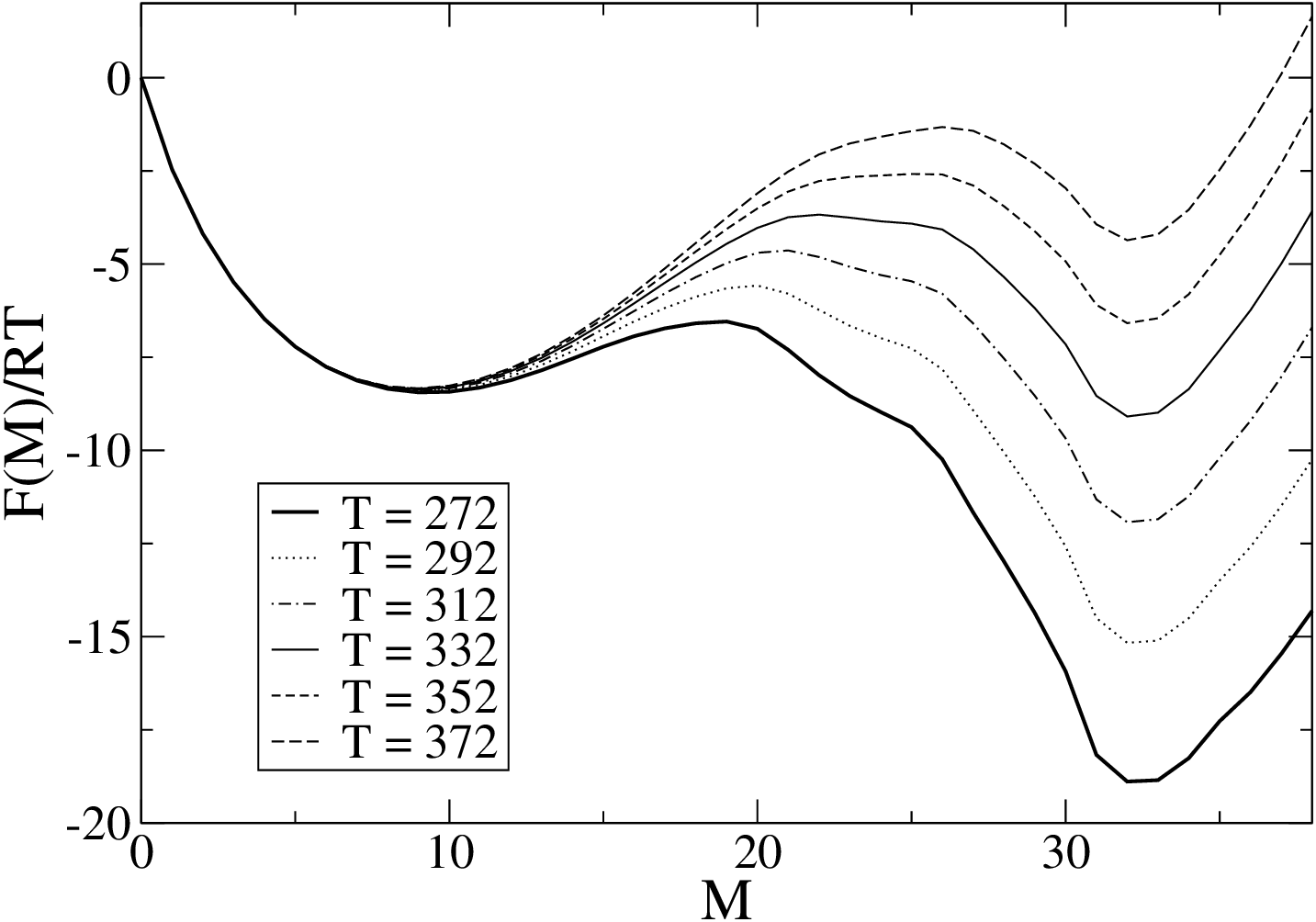}}
\caption{\label{fig:EqLandPIN1} The PIN1 free energy profile
  $F(M)$ for various temperatures.}
\end{figure}

For comparison, we report the same quantity for the WW-domain of protein PIN1 in Fig.\
\ref{fig:EqLandPIN1}. One can see that both molecules can exhibit a
barrier in their profile, though 1BBL has a much smaller barrier,
which indeed disappears at low enough temperature, and a much wider
unfolded minimum with respect to PIN1, whose free energy profile
exhibits a well definite two--state structure. At $T = 289$ K, the wide
plateau--like region in the range $M \sim 10$ to 25 for 1BBL corresponds
approximately to the wide fluctuations observed in the single MC
trajectory. 
Notice that upon raising the  temperature this flat region moves towards more unfolded states, and this shift  gives rise to the long tail we can see in Fig.\ \ref{fig:pFit}, while a barrier appears in the vicinity of the native state and then move towards it.
On the other end, at
$T = 269$ K, one gets an almost linear profile,
decreasing with $M$. In order to get a similar profile for PIN1 one
should go to much lower temperatures. It is not possible to define a
sharp threshold, but the value of 170 K obtained in the previous
section is certainly a good estimate. 

It is also interesting to compare the barrier heights observed above
with the rates computed in the previous section. 

\begin{figure}
\centerline{\includegraphics*[width=8cm
]{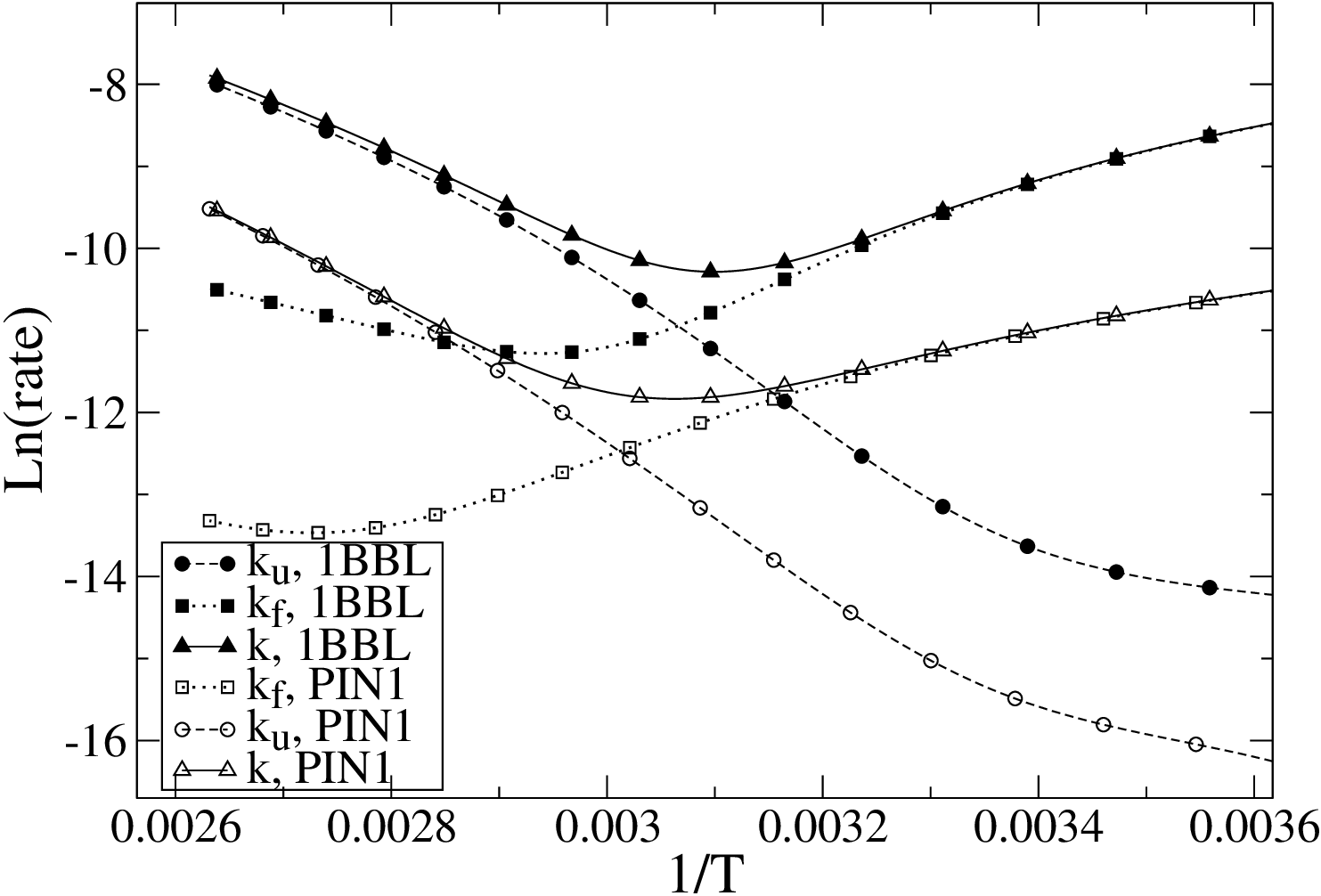}}
\caption{ Natural logarithm of the folding, unfolding and relaxation rates of protein 1BBL and of the WW-domain of PIN1, vs 1/T. Rates are measured in units of the inverse of the elementary time, corrisponding to flipping the state of one peptide-bond. Temperatures are in Kelvin. }
\label{fig:arrheniusplot}
\end{figure}

Indeed, if (and only if) we assume a two-state behavior, it is possible to define folding and unfolding rates $k_f$ and $k_u$, that are necessarily related to the measured equilibration rate $k$ and average fraction of native peptide units $m$ by the equations $k=k_f+k_u$ and  $k_f/k_u = p/(1-p)$.
Figure \ref{fig:arrheniusplot} reports the results for the folding, unfolding and global equilibration rate for 1BBL and PIN1: it can be noticed that values of $k_f$ increasing with temperature appear in both proteins, signalling the region were the two-state assumption fails.
However, for the WW-domain  this happen just at temperatures greater than 367 K, well above the mid-point of the transition, while for 1BBL $k_f$ presents a minimum at T=340 K, rather close to $T_m$=326.2 K. 
Then, assuming the
Arrhenius--like behavior involved by the two-state hypothesis, we looked for a linear correlation between
the natural logarithm of the folding rate $\ln k_f$ and the height of
the folding barrier divided by $RT$. 
%
We find that the linear relationship between $\ln k_f$ and $\Delta F_{\dagger - U}/(R T)$ (where $\dagger$ indicates the barrier top and $U$ the unfolded minimum) exists only for   values of the barrier below 3.8 R T approximately (corresponding to the region of temperatures 302 -- 340 K); then the folding rate would present a minimum and would start to increase for increasing barrier, signalling that the two-state hypothesis cannot be applied any more. 
%
For PIN1, we find that the linear correlation is  very good in the larger range of temperatures 296-350 K.
The linear fits reveal that for 1BBL the absolute value of the slope is 0.6, while for PIN1 it is 0.85 (data not reported), which is again an indication of the bigger deviation of 1BBL from two-state behavior. The fact that the slope is not 1 even for PIN1 is probably related to the above-mentioned small overestimation of the rates by LEA.
%
%
%
In summary, based on the above results,
 one cannot regard 1BBL as a true downhill
folder, since the typical features of a downhill folder appear only at
low enough temperatures, and not in the whole temperature range. It is
however true that its behavior differs qualitatively from that of PIN1 WW-domain, since
to get a similar behavior for the latter one would have to go down to
unphysical temperatures.

\section{Conclusions} 
\label{sec:concl}

We have analyzed the folding behavior of protein 1BBL, which has been
proposed as a downhill folder, compared to that of the WW-domain of protein PIN1, which is a
two--state folder, in the framework of the WS model, a
statistical mechanical G\={o}--like model with binary variables. The
model has been solved exactly at equilibrium, while for the kinetics
we have used Monte Carlo simulations and the semi--analytical local
equilibrium approach. The model parameters have been determined by
means of an analysis which indicates that the distinction
which is usually made, between the entropic contributions of the more
structured and the less structured parts of the chain, is not crucial in determining the overall behavior of the system. This result allows to reduce the number of parameters and
can be useful also in future studies based on the same model. 

Comparing the behavior of the two molecules we have seen that 1BBL
departs from  a clear two--state behavior 
in several respects. 
Deviations from a single exponential behavior are more
relevant than in the case of PIN1, single MC trajectories can differ
markedly from those of a two--state folder at low enough temperature,
and free energy profiles exhibit a very small barrier, which vanishes
upon lowering the temperature. All these features would indeed agree
with the usual expectations for a downhill folder if they were
exhibited in a wide temperature range, and not just at very low
temperature. However, we do find a small barrier in the free-energy profile around the temperature of the transition midpoint $T_m$, as measured by half  the variation of the fraction of folded residues between the two asymptotic values of low and high temperatures (i.e. the native and unfolded baselines).  This prevents us from considering 1BBL as a true downhill folder, and suggests that the picture of a fast, but barrier-limited, folding applies at least  in the region around $T_m$. 
On top of this, one should consider that, at this extremely simplified level of description, our model appears to be less cooperative than the real protein, as Fig.\ \ref{fig:pFit} suggests, so that the similarity to a two-state folder could be also more pronounced in reality. 

One could, of course, argue that even PIN1 can exhibit the same
phenomena as 1BBL, albeit at unphysically low temperatures. Indeed, as it has
been already recognized in the literature, two--state and downhill
folders do not make up two well separate classes, they are just two
extreme cases of a range of possible behaviors. It is however true
that the crossover between these two different behaviors occurs, at
least in the present model, in a temperature range which is
not biologically relevant for PIN1, while it is close to physiological
for 1BBL. Therefore, our results can be regarded as a confirmation that
1BBL cannot either be considered a clear two--state folder. It is rather an
example of a molecule which crosses over from a weak two--state
behavior at the denaturation temperature to a downhill folding
behavior at lower temperatures.

Our results are apparently
at odds with the theoretical results reported by Wang and coworkers \cite{wangprot2006,wangcpl2005}, that use Langevin dynamics of a G\={o}--like model to study 1BBL, and find that the downhill scenario holds at all temperatures within their modelling scheme. However, as they point out, different folding behavior appears  to be triggered by the average number of non-local contacts per residue, and this could explain the differences  between their and our results, since we resort to a different definition of the contact map, with all atom-atom contacts, since it allows a better fit of the parameters of the WS model to the experimental data. On the other hand, we find that this choice increases the ratio of non-local to local contacts: we have that the average number of non-local contacts per residue is $N_N=1.65$, which is above the threshold that Wang and coworkers report for two-state behavior.
This also propose, indeed, a word of caution about the model-dependent features of theoretical results, along the very same lines of the results reported in \cite{Rey}, where an analysis is performed of the influence of the cutoff distance, used to define the contact map, on the degree of cooperativity of the folding transition.
The robust features that appear from our results are that protein 1BBL is indeed less cooperative that the WW-domain of protein PIN1, and departs significantly from a clear two-state behavior. 
\begin{acknowledgments}
P.B. acknowledges financial support from MEC (Spanish Education and Science Ministry) through research contracts  FIS2004-05073 and FIS2006-12781. 
P.B. is also grateful to Adri\'an Vel\'azquez Campoy for fruitful discussions.

\end{acknowledgments}


\end{document}